\documentclass[twocolumn,aps,showpacs,preprintnumbers,amsmath,amssymb]{revtex4}
\usepackage{graphicx}

\begin{document}

\title{ Ultraslow light in inhomogeneously broadened media}

\author{G. S. Agarwal}
\altaffiliation{On leave of absence from Physical Research
Laboratory, Navrangpura, Ahmedabad - 380 009, India.}
\author{T. N. Dey}

\affiliation{Department of Physics, Oklahoma State University,
Stillwater, OK - 74078, USA}
\date{\today}

\begin{abstract}
We calculate the characteristics of ultraslow light in an
inhomogeneously broadened medium. We present analytical and
numerical results for the group delay as a function of power of
the propagating pulse. We apply these results to explain the
recently reported saturation behavior [Baldit {\it et al.}, \prl
{\bf 95}, 143601 (2005)] of ultraslow light in rare earth ion
doped crystal.
\end{abstract}


\maketitle

 The usage of a coherent field to control the optical properties of
 a medium has led to many remarkable results such as enhanced
 nonlinear optical effects \cite{Tewari_PRL_86,Harris_PRL_90},
 electromagnetically induced transparency (EIT)
 \cite{Harris_PT_97}, lasing without inversion
 \cite{Kocharovskaya_JETP_88,Kocharovskaya_PR_92,Agarwal_PRA_91},
 ultraslow light \cite{Hau_Nat_99,Kash_PRL_1999,Budker_PRL_99,Turukhin_PRL_2002,Ku_OL_04},
 storage and retrieval of optical pulses \cite{Lukin_PRL_01} and many
 others \cite{Zibrov_PRL_96,Kang_PRL_04,Deng_PRL_03,Stenner_Nature_2003}. Most of these
 effects rely on quantum interferences which are created by the
 application of a coherent field. The coherent field opens up a new
 channel for the process under consideration. This interference
 effect produces the EIT dip or a hole in the absorption profile.
 The ultraslow light emerges as the EIT dip could be very narrow.
 It has been realized that in principle one could also use two
 level nonlinearities in presence of a strong pump. For a
 homogeneously broadened medium a hole can emerge if the transverse
 and longitudinal relaxation times are quite different. Under these
 conditions the hole has a width of the order of $T_1$ and this is
 being referred to as the effect of coherent population
 oscillation \cite{Schwartz_APL_67}. Bigelow {\it et al.} did experiments in this regime using
 ruby as the material medium
 which can be modelled as a homogeneously broadened system \cite{Bigelow_PRL_2003,Bigelow_SCI_2003}.
 Some studies on slowlight in inhomogeneous broadened medium exist \cite{Kocharovskaya_prl_01,Agarwal_PRA_03}. In an earlier paper the
 present authors had considered the case of inhomogeneously
 broadened gaseous medium where the Doppler effect is
 important \cite{Agarwal_PRA_03}. We considered the case of
 saturation absorption spectroscopy. This leads to the well known
 hole in the Doppler profile. The width of this hole was of the
 order of $1/T_1$ which is about two times $1/T_2$. In the
 inhomogeneously broadened gaseous medium the group index of the
 order of $10^3$ was obtained. The recent experiment of Baldit {\it et al.}
 reports group delays of the order of 1.1 s in rare earth ion
 doped crystal which has strong inhomogeneous broadening \cite{Baldit_PRL_05}. In this case all the
 relaxation times are quite different - $T_1= 8$ ms; $T_2=3~\mu$s;
 inhomogeneous line width $\Gamma_{inh}=1.3$ GHz. The width of the whole is
 essentially determined by $T_1$ and hence one gets very large
 delays. Baldit {\it et al.} did present a theoretical model based on
 homogeneous broadening of the medium where as to obtain agreement
 with experiments inhomogeneous broadening is to be included as alluded by them \cite{footnote1}.

 In this paper, we consider  a system of inhomogeneously broadened
 two level atoms interacting with co-propagating pump and probe
 fields. We use the well known susceptibility
 \cite{Mollow_PRA_1972} and average it over the inhomogeneous
 distribution to calculate the group index. We derive a number of
 analytical results and show how these can be used to understand
 the experimental results of Baldit {\it et al}. For example we
 show that in the limit of very small detuning of the probe from
 the pump the group delay goes as $\sqrt{S}$ for large S. The group
 delay also peaks at about S=0.9. The value of group delay increases
 as the detuning $\delta$ increases. We further present detailed numerical
 results.

\begin{figure}
 \scalebox{0.55}{\includegraphics{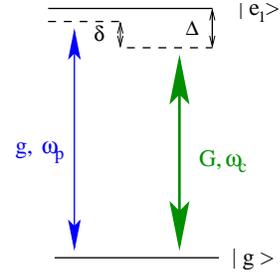}}
 \caption{\label{Fig1} A schematic diagram of two level atomic
 system with ground state $|g\rangle$ and excited state
 $|e_1\rangle$; The pump $(\omega_c)$ and probe $(\omega_p)$
 fields are co-propagating.}
 \end{figure}

\begin{figure*}
 \begin{tabular}{cc}
 \scalebox{0.45}{\includegraphics{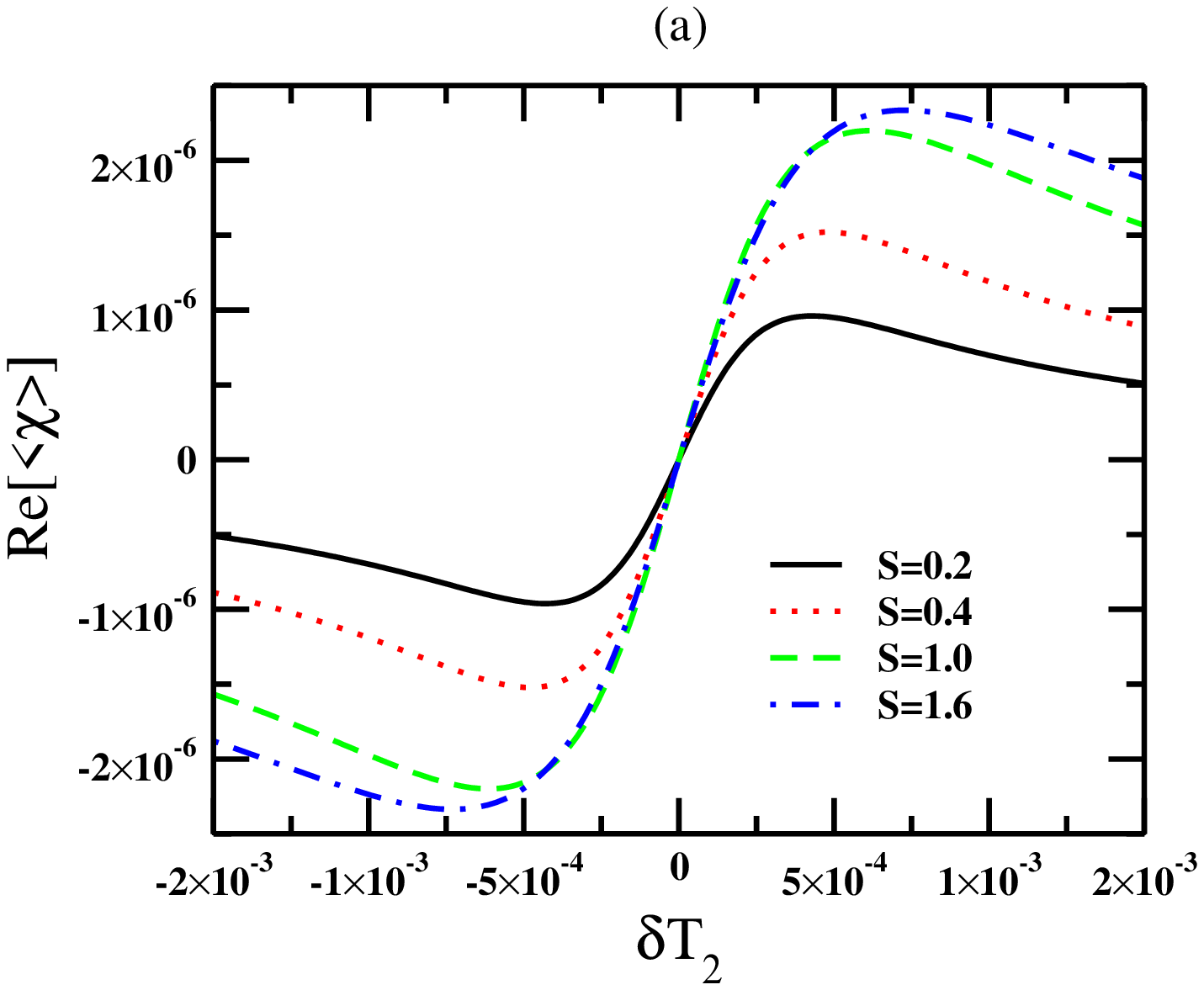}}&
 \scalebox{0.45}{\includegraphics{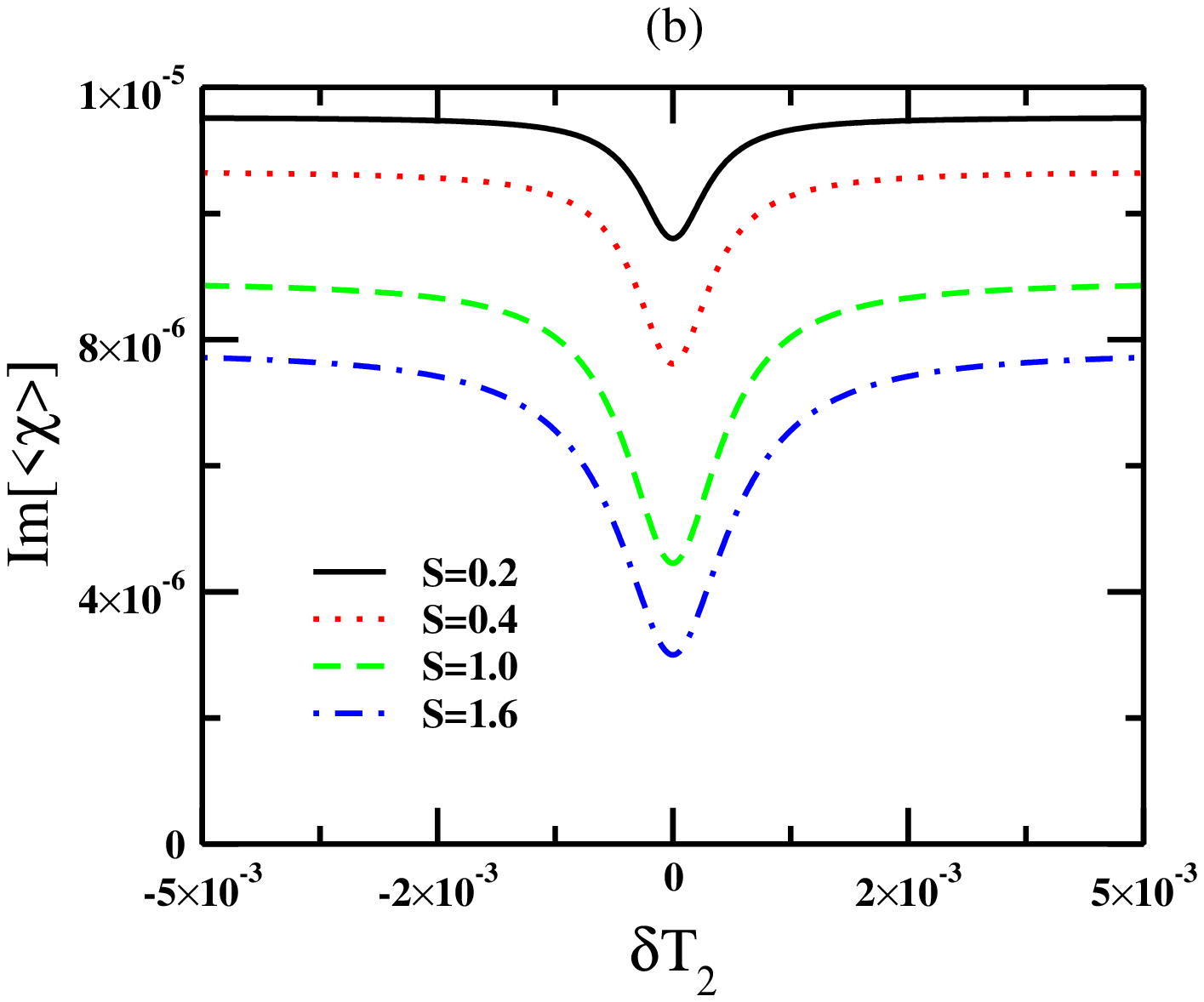}}
 \end{tabular}
 \caption{\label{Fig2} (a) and (b) The real and imaginary parts respectively
 of the susceptibility $\langle\chi\rangle$ at the probe frequency $\omega_p$ in the presence
 of co-propagating control field G. The common parameters of the five plots
 for erbium doped crystal are chosen as: inhomogeneous absorption coefficient
 $\alpha_{inh}=6.5$ cm$^{-1}$; inhomogeneous line width $\Gamma_{inh}=1.3$ GHz;
 longitudinal relaxation time T$_1$= 8 ms;
 transverse relaxation time T$_2$=3 $\mu$s. }
 \end{figure*}

 In order to understand the experimental results of Baldit {\it et
 al}, we consider a two level system as shown in figure~\ref{Fig1}.
 Here we define all fields as
 \begin{equation}\label{field}
 \vec{E_i}(z,t)=\vec{\mathcal E_i}(z,t)~e^{-i(\omega_i
 t-kz)}~+~c.c.,(i=p,c)
 \end{equation}
 where $\vec{\mathcal E_i}$ is the slowly varying envelope of the
 field. The pump field at frequency $\omega_c$ and the probe field at
 frequency $\omega_p$ are co-propagating through the medium. The
 linear susceptibility $\chi(\omega_p)$ is obtained by solving the
 density matrix equations for the two level system of
 figure~\ref{Fig1}, that is by calculating the density matrix
 element $\rho_{eg}$ to the first order in the probe field but to
 all orders in the co-propagating pump field. The dynamics of
 population and polarization of the atoms in the
 two-level configuration are given by
 \begin{eqnarray}\label{density_matrix}
 &\dot{\rho}_{_{ee}}&=-\frac{1}{T_{1}}\rho_{_{ee}}
 +i\left(G + ge^{-i\delta t}\right)\rho_{_{ge}}-i\left(G^* + g^*e^{i\delta
 t}\right)\rho_{_{eg}},\nonumber\\
 &\dot{\rho}_{_{eg}}&=-\left(i\Delta+\frac{1}{T_{2}}\right)\rho_{_{eg}}
 +i\left(G + ge^{-i\delta
 t}\right)(\rho_{_{gg}}-\rho_{_{ee}}),\nonumber\\
 &\rho_{_{ee}}& + \rho_{_{gg}}=1
 \end{eqnarray}
 where $T_1$ and $T_2$ are the longitudinal and transverse relaxation times respectively.
 The density-matrix elements in the original frame are given by
 $\rho_{_{eg}}e^{-i\omega_c t},\rho_{_{gg}}$, and $\rho_{_{ee}}$.
 The detunings $\Delta,\delta$ and the Rabi frequencies are
 defined by
 \begin{equation}
 \Delta=\omega_{eg}-\omega_c;\delta=\omega_p-\omega_{c};
 2G=\frac{2\vec{d}_{eg}.\vec{\cal E}_c}{\hbar};2g=\frac{2\vec{d}_{eg}.\vec{\cal
 E}_p}{\hbar},
 \end{equation}
 where $\vec{d}_{eg}$ is the dipole matrix element.
 The susceptibility $\chi$ can be obtained by considering the
 steady state solution of Eq.(\ref{density_matrix}) to the first
 order of $g$ and write the solution as
 \begin{equation}\label{peturbation}
 \rho=\rho^0~+~g~e^{-i\delta
 t}~\rho^+~+~g^*~e^{i\delta t}~\rho^-+......
 \end{equation}
 The eg element of $\rho^{+}$ will yield the linear susceptibility $\chi$ at the frequency $\omega_p$
 as can be seen by combining Eqs.(\ref{density_matrix}) and
 (\ref{peturbation}):
 \begin{widetext}
 \begin{equation}
 \chi=-\frac{n|d|^2T_2}{\hbar}\frac{1+\Delta^2{T_2}^2}
 {(1+\Delta^2{T_2}^2+S)(\Delta T_2+\delta T_2+i)}
 \left[1-\frac{S(\Delta T_2-i)^{-1}(\delta T_2+2i)(\delta T_2-\Delta T_2+i)}
 {2(\delta T_1+i)(\delta T_2+\Delta T_2+i)(\delta T_2-\Delta T_2+i)-S(\delta T_2+i)}\right],
 \end{equation}
 \end{widetext}
 where $n$ is the density of the
 atoms of the medium. The saturation parameter $S=4|G|^2T_1T_2$ is defined as the ratio of the
 control field intensity and the saturation intensity. The average response of
 the susceptibility is given by
 \begin{equation}
 \langle\chi\rangle=\frac{2\sqrt{ln2}}{\sqrt{\pi}\Gamma_{inh}}\int\chi(\Delta) e^{
 -\frac{4\ln 2 \left[\Delta-(\overline{\omega}_{eg}-\omega_c)\right]^2}{\Gamma_{inh}^2}} d
 \Delta,
 \end{equation}
 where $\overline{\omega}_{eg}$ is the central frequency of the atomic
 transition $|e\rangle \longleftrightarrow |g\rangle$. Here we consider
 the frequency of the control field $\omega_c$ is tuned to the line center $\overline{\omega}_{eg}$.
 We present the
 behavior of real and imaginary parts of the susceptibility as a
 function of the detuning of the probe field in
 Fig.~(\ref{Fig2}). The real part of susceptibility gives normal
 dispersion. It is clear from Fig.~(\ref{Fig2}a) that
 the slope of normal dispersion attains maximum when S $\sim 1$ which leads
 to ultra slow light. The imaginary part of $\langle\chi\rangle$
 exhibits the absorption dip which becomes deeper with the
 increase in the intensity of the control field as shown in Fig.~(\ref{Fig2}b). The spectral width of
 absorption dip depends on the intensity of the control field.
 This dip is associated with coherent population
 oscillation\cite{Schwartz_APL_67}.

 In order to compare with experimental results of Baldit {\it et al.}
 we need to know the group index $n_g$ which is defined by
 \begin{eqnarray}\label{group_index}
 n_g &=& 1 + 2\pi \omega_p \frac{\partial}{\partial \omega_p}{\rm
 Re}\langle\chi\rangle\nonumber\\
 &=& 1-
 \frac{\alpha_{{\tiny inh}} cT_2}{2\pi}
 \left\langle D
 \right\rangle
 \end{eqnarray}
 where
 \begin{widetext}
 \begin{equation}
 D=
 \frac{
 i(\Delta+i)
 \left[S^2+2(\delta f+i)^2(i+\delta -\Delta)^2(1+i\Delta)
 +S(i+\delta-\Delta)\left(-i+\delta+2f(i+\delta-i\delta^2-\Delta)+\Delta\right)
 \right]}
 {
 2
 \left(1+S+\Delta^2\right)
 \left[
 S(\delta +i)-(i+\delta f)
 \left(
 (i+\delta)^2-\Delta^2
 \right)\right]^2
 };~f= \frac{T_1}{T_2}.
 \end{equation}
 \end{widetext}
 We denote the integration with respect to $\Delta$ has been denoted by $\langle\rangle$.
 The unsaturated inhomogeneous absorption coefficient
 of the two level atomic system is defined as
 \begin{equation}
 \alpha_{inh}=\frac{4\pi\omega_p}{c}\langle Im\left[\chi\right]_{_{G=0}}
 \rangle=\frac{8\pi^{\frac{3}{2}}\omega_p n |d_{eg}|^2 \sqrt{\ln 2}}{c\hbar\Gamma_{inh}}
 \end{equation}
 In the limit of very small detuning of the probe from
 the pump, the analytical expression
 for the group index for
 inhomogeneous case can be expressed as
 \begin{equation}\label{index_inhana}
 n_g\cong
 c\alpha_{inh}T_1\left[\frac{S(4+S)}{16(1+S)^{5/2}}\right];~~\delta \rightarrow 0
 \end{equation}
 It is clear from the above expression that the group index varies as $S^{-1/2}$
 for large value of $S$. The group index attains the maximum value at
 $S=0.9$. In case of homogeneously broadened two level system the group index
 is given by \cite{Baldit_PRL_05}
 \begin{equation}\label{index_hana}
 n_g\cong c\alpha_{h}T_1\left[\frac{S}{2(1+S)^{3}}\right];~~\delta \rightarrow
 0,
 \end{equation}
 where $\alpha_{h} = 4\pi\omega_p n |d_{eg}|^2 T_2/c\hbar$ is the homogeneous absorption coefficient. For homogeneous
 two level system the group index varies as $S^{-2}$ at large S
 and peaks at S=0.5.
 \begin{figure}
 \scalebox{0.65}{\includegraphics{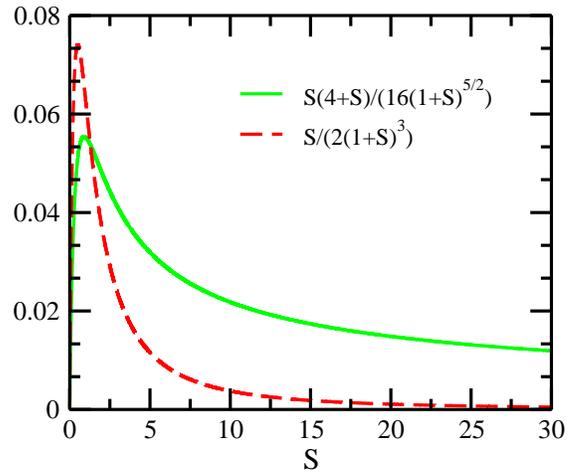}}
 \caption{\label{Fig3}Variation of the term in squared bracket of
 Eqs.~(\ref{index_inhana}) and (\ref{index_hana}) as a function of
 intensity of the control field for inhomogeneous and
 homogeneous cases of two level system.}
 \end{figure}
 At large S, the group index for a two level system falls much slowly for
 an inhomogeneous medium as compared to the homogeneous case as shown in Fig.~(\ref{Fig3}).
 We thus find an important difference between inhomogeneously and
 homogeneously broadened two level systems.
 Note that the ratio between
 inhomogeneous and homogeneous unsaturated absorption
 coefficient is $\alpha_{inh}/\alpha_{h}\approx \Gamma_{inh}T_2$.
 The behavior so obtained is consistent with the experimental observation.
 Figure~(\ref{Fig4}) shows the variation of group index as
 function of the intensity of the control field at different probe
 detuning.
\begin{figure}
 \scalebox{0.6}{\includegraphics{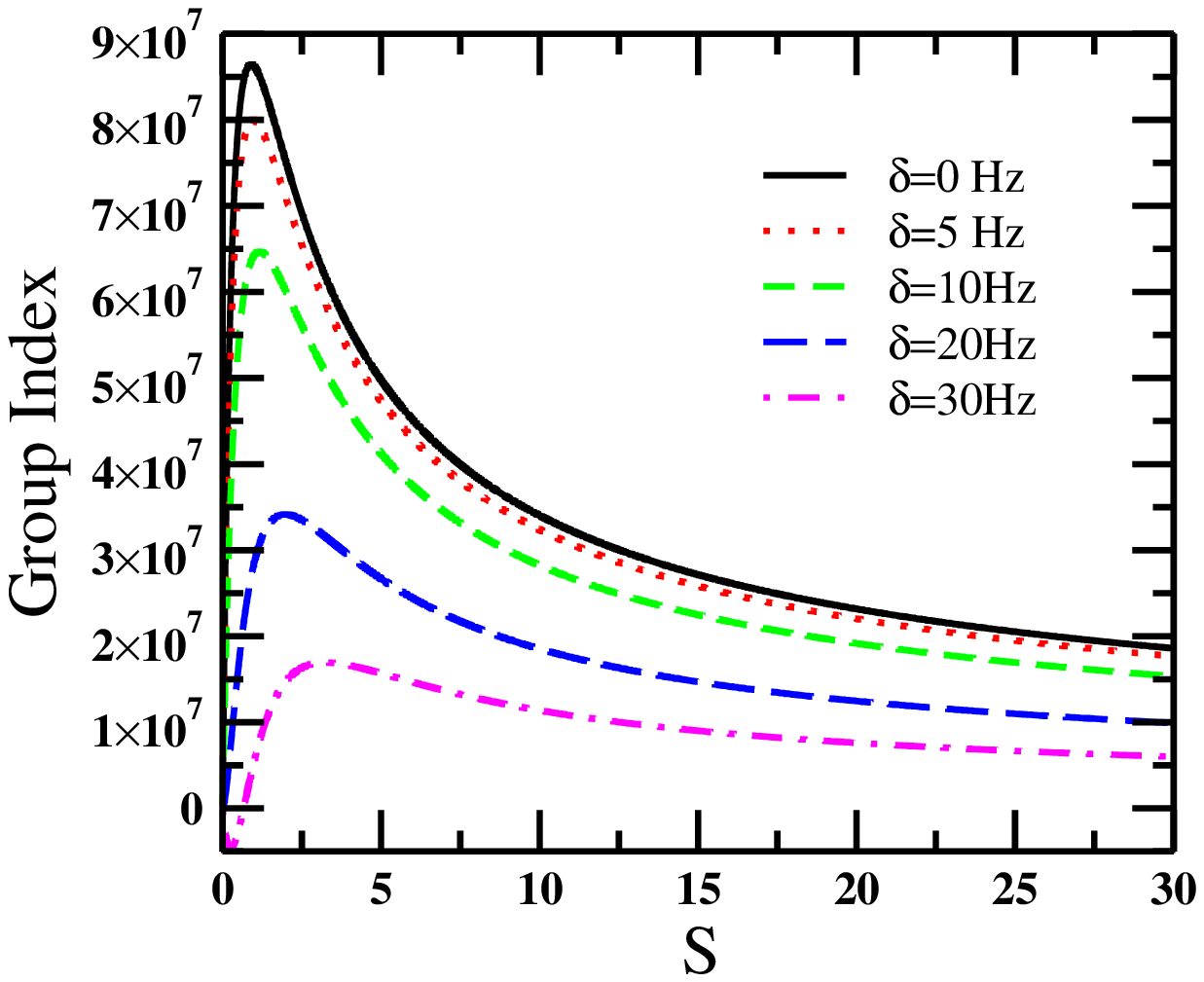}}
 \caption{\label{Fig4} The variation of group index with the saturation
 parameter S. The parameters are chosen as $\alpha_{inh}$=6.5 cm$^{-1}$, T$_1$=8 ms and
 T$_2$= 3 $\mu$s.}
 \end{figure}
 As the detuning of the probe field is increased the peak of the
 group index shifts toward higher S. The maxima of the group index $n_g=.65\times
 10^{8}$ for $\delta=10$ Hz occurs at $S=1.16$  which corresponds to the group velocity
 $v_g=c/n_g=4.61$ m/s which is higher than what is reported. Further the
 figure~(\ref{Fig5}) shows the variation of the group index
 $n_g$ calculated from the Eq.~(\ref{group_index}) with the
 detuning of the probe field $\delta$ for saturation parameter
 S=1.

 \begin{figure}
 \scalebox{0.62}{\includegraphics{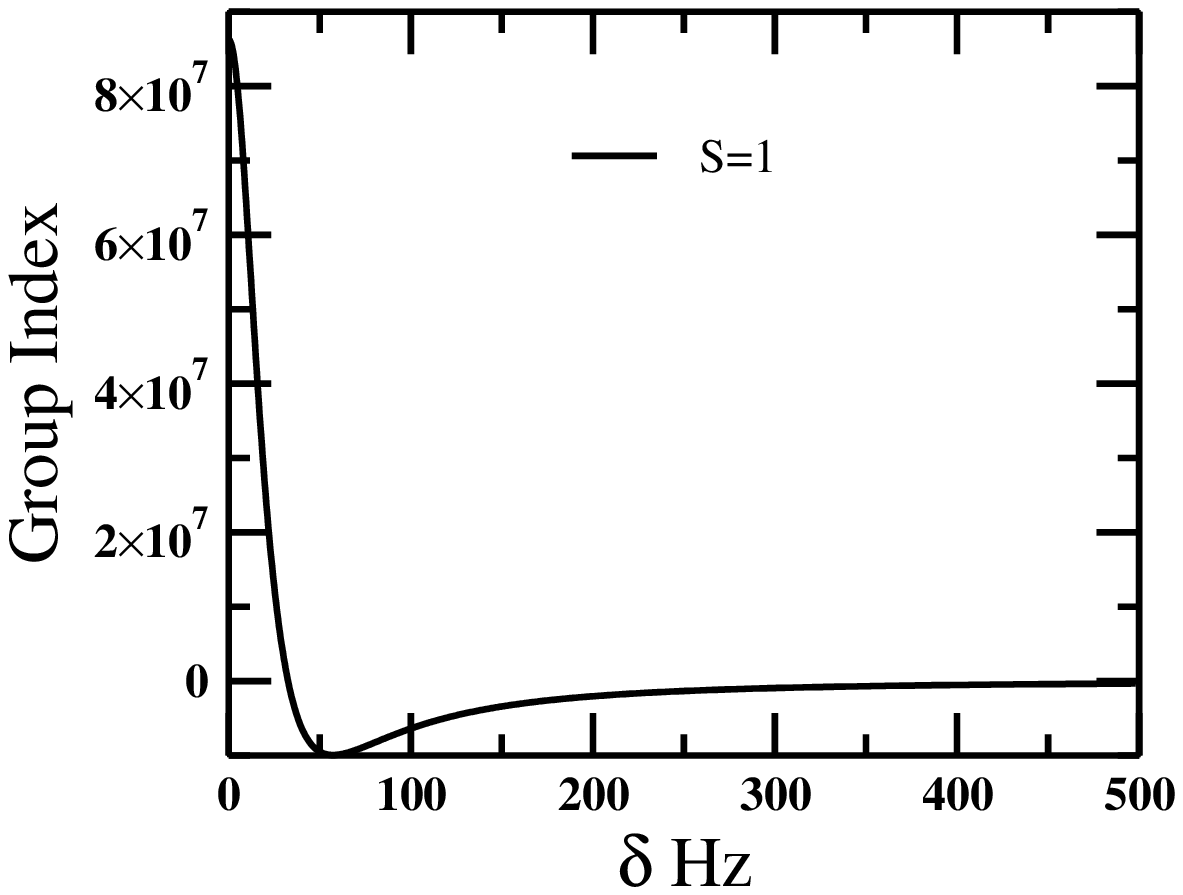}}
 \caption{\label{Fig5} Group index variation with the detuning of the probe field.
 The different parameters used in the numerical simulation of the
 Eq.~(\ref{group_index}) are as follows: $\alpha_{inh}$=6.5 cm$^{-1}$, T$_1$=8 ms and
 T$_2$= 3 $\mu$s.}
 \end{figure}

 In conclusion, we have discussed the characteristics of ultraslow
 light in an inhomogeneously broadened medium. Our numerical and
 analytical results enable to understand the nature of the experimental results
 of Baldit {\it et al}. However the results derived here are applicable
 to any system which can be modelled by inhomogeneously broadened
 two level system.


\begin{thebibliography}{999}
\bibitem{footnote1}There are important
 differences between the Doppler broadened gaseous medium and solid
 state medium like rare earth ion doped crystal. These differences
 arise from the way the susceptibilities are to be averaged over
 inhomogeneous distribution.
\end{thebibliography}
 \end{document}